\documentstyle[twocolumn,prl,aps,psfig]{revtex}
\begin{document}

\title{Drift without flux: Brownian walker with a space dependent
diffusion coefficient}

\author{P. Lan{\c c}on$^1$ ,  G. Batrouni$^2$ , L. Lobry$^1$  and
N. Ostrowsky$^1$}

\address{$^1$
Laboratoire de Physique de La Mati\`ere Condens\'ee, CNRS UMR 6622,
Universit\'e de Nice--Sophia Antipolis,  
Parc Valrose, 06108 Nice, France}
\address{$^2$
Institut Non-Lin\'eaire de Nice, CNRS UMR 6618,
Universit\'e de Nice--Sophia Antipolis, 
1361 route des Lucioles,
06560 Valbonne, France}
\date{\today}
\address{\mbox{ }}        

\address{\parbox{14cm}{\rm \mbox{ }\mbox{ }     
Space dependent diffusion of micrometer sized particles has been
directly observed using digital video microscopy. The particles were
trapped between two nearly parallel walls making their confinement
position dependent.  Consequently, not only did we measure a diffusion
coefficient which depended on the particles' position, but also report
and explain a new effect: a drift of the particles' individual
positions in the direction of the diffusion coefficient gradient, in
the absence of any external force or concentration gradient.}}
\address{\mbox{ }}
\address{\parbox{14cm}{\rm \mbox{ }\mbox{ }
PACS number(s): 05.40.Jc, 82.70.Dd, 67.40.Hf}}
\address{\mbox{ }}
\maketitle
\narrowtext  
\vskip 0.7cm

Brownian motion of spherical colloidal particles in the vicinity of a
wall has been extensively studied, both theoretically and
experimentally . It has been shown that the diffusion coefficients
parallel or perpendicular to the wall were greatly reduced when the
particles were close enough to the obstacle, i.e. within distances
comparable to or less than their radius\cite{happel}. When the Brownian
particles are trapped in a more confined geometry, such as a porous
medium, the theory is far more complicated and few experimental
studies have been reported in model geometries, where the particles
are trapped between two parallel walls\cite{fauch,carba}. In this article, we
report some new experimental results concerning the Brownian motion of
particles trapped between two {\it nearly} parallel walls, so that the
confinement, and thus the diffusion coefficient, become space
dependent. As a result, we not only measure a diffusion coefficient
which varies with the confinement, but also a drift of the particules'
individual positions in the direction of the diffusion coefficient
gradient, in the absence of any external force or concentration
gradient. This drift was not accompanied by any net particle flux,
i.e. statistically the same number of particles crossed any imaginary
surface in both directions.  We first discuss the general problem of a
Brownian walker with a spatially dependent diffusion coefficient to
explain the origin of the expected drift, and then present the
experimental set-up and results.  

As in our experiment the diffusion
coefficient varies in only one direction, say $x$, we briefly sketch a
heuristic derivation of the 1D Brownian walker algorithm. The velocity
of a 1D Brownian particle subjected to a random force and a viscous
drag follows the Langevin equation,
\begin{equation} 
{dv(t) \over dt} = -\gamma v(t) +\Gamma(t),
\label{lang-v}
\end{equation} 
where $\gamma^{-1}$ is the velocity relaxation time and $\Gamma(t)$
the random force per unit mass defined by its mean value $\langle
\Gamma(t)\rangle=0$ and correlation function $\langle
\Gamma(t)\Gamma(t^{\prime})\rangle=q \delta(t-t^{\prime})$.  Using the
equipartition theorem it can be shown that $q$ is related to the
temperature $T$ and the particle's mass, $m$, by the standard relation
$q=2\gamma kT/m$. Discretizing the random function $\Gamma(t)$ over
time intervals $\Delta t>>\gamma^{-1}$ allows us to drop in Eq.(1) the
inertial term, ${dv/dt}$, and to replace the velocity $v$ by
$\Delta x/\Delta t$.  Choosing for $\Gamma(t)$ the simplest random
function, $\Gamma(t)=\pm \sqrt{q/\Delta t}$, leads to the well known
Brownian walker algorithm,
\begin{equation} 
x(t+\Delta t) = x(t) \pm \sqrt{2D\Delta t}
\label{walker}
\end{equation} 
with $D=kT/m\gamma$. When the diffusion coefficient $D$, i.e. when the
temperature $T$ and/or the drag coefficient $\gamma$ become position
dependent, the above algorithm needs to be clarified.  During each
time interval $\Delta t$, the walker makes a step to the right or to
the left, but should the length of this {\it position dependent} step,
$\sqrt{2D\Delta t}$, be computed at the departure point $x(t)=x$, the
arrival point $x(t+\Delta t)=x+\Delta x$ or at any point in between?
These mathematical choices, often referred to as the Ito/Stratonovitch
conventions\cite{risken}, {\it model different physical situations and the
choice of convention is dictated only by the physics}.  We denote by
$D(x+\alpha \Delta x)$ the diffusion coefficient appearing in Eq.(2)
where $\alpha =0,1/2$ and $1$ correspond to the Ito, Stratonovitch
and {\it isothermal} choices respectively. As we will show, this last
case models a situation where the temperature, $T$, is uniform but the 
drag coefficient, $\gamma$, is space dependent. Using in Eq.(2) the 
limited expansion,
$D(x+\alpha \Delta x) \approx D(x)+\alpha(dD/dx)\Delta x$ with $\Delta
x=\pm \sqrt{2D(x)\Delta t}$, yields the algorithm for a Brownian
walker with a position dependent diffusion coefficient:
\begin{equation} 
x(t+\Delta t) = x(t) \pm \sqrt{2D(x(t))\Delta t}+\alpha {{dD}\over
{dx}}\Delta t.
\label{walker2}
\end{equation}
Depending on the value of $\alpha$, this model has very different
implications concerning the equilibrium distribution of the Brownian
walkers, their individual drift($\langle x(t)-x(0)\rangle$) and their
net flux.

Averaging Eq.(3) over a large number of walkers shows the average
position of a Brownian walker is no longer zero since the diffusion
gradient term acts as an external force leading to particle drift. If
this gradient is assumed to be constant, this drift increases linearly
with time as
\begin{equation} 
\langle x(t)-x(0)\rangle = \alpha {{dD}\over {dx}}\Delta t.
\label{drift}
\end{equation}
A first intuitive, but misleading, idea would be to conclude that the
particles will migrate in the direction of the diffusion gradient,
leading, therefore, to a concentration gradient.  This is actually
incorrect as we now show. Starting with a uniform particle
distribution $\rho_0$, we check if this corresponds to an equilibrium
state by determining the particle flux through an imaginary surface,
$S$, placed perpendicular to the diffusion coefficient gradient at
coordinate $x$ (see Fig. 1). During a time interval $\Delta t$, all
particles crossing $S$ from the left (or right) are half of those
included in the volume $S L_{right}$ ($S L_{left}$) where $L_{right}$
($L_{left}$) is the right (left) step {\it terminating at} $x$ taken 
by a walker during $\Delta t$. The net particle flux to the right will
thus be
\begin{equation} 
J = {{\rho_0}\over {2}} {{S L_{right}-S L_{left}}\over{S\Delta t}}.
\label{flux}
\end{equation}
Equation (2) allows computing the length of these two steps which 
both end at the same point x:
\begin{equation} 
L_{^{right}_{left}} = \sqrt{2D(x)\Delta t} \pm (\alpha -1){{dD}\over
{dx}}\Delta t,
\label{steplength}
\end{equation}
leading to the particle flux
\begin{equation} 
J = -{\rho_0} (1-\alpha) {{dD}\over {dx}}.
\label{flux2}
\end{equation}

\begin{figure}
\psfig{file=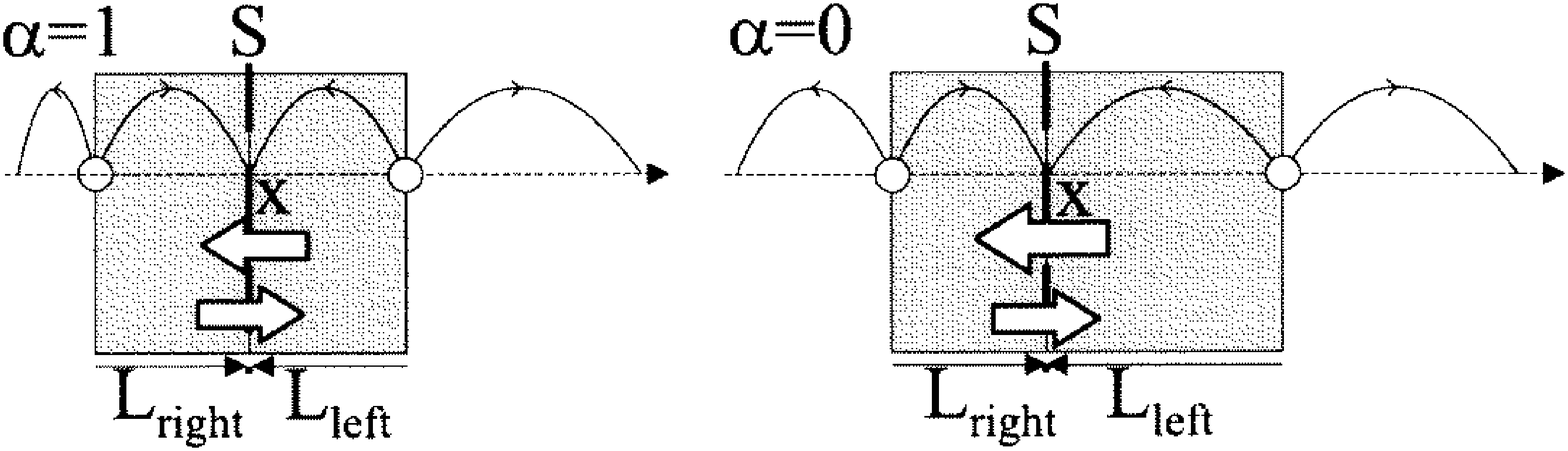,height=2.0in,width=3.0in}
\vskip-00mm
\caption{ 
 Particle flux for Brownian walkers with a step length depending
on arrival position ($\alpha=1$) or departure position ($\alpha=0$).
}
\end{figure}

As a result, in the situation of maximum drift where $\alpha=1$, this
flux will vanish (see left part of Fig. 1), meaning that the uniform
particle distribution corresponds to an equilibrium. According to
Boltzmann, this should correspond to an isothermal situation, the
diffusion coefficient gradient arising only from a pure hydrodynamic
effect, the spatial dependence of the drag coefficient $\gamma$. For
all the other values of $\alpha$, the flux will be negative, leading
to a concentration gradient of the particles in the direction opposite
to that of the diffusion coefficient gradient.  The maximum flux (and
zero drift) is obtained for $\alpha=0$, as shown on the right part of
figure 1. Equation (7) may be generalized to the case where the
particle distribution is position dependent:

\begin{equation} 
J = -(1-\alpha)\rho(x){{dD}\over {dx}}-D{{d\rho}\over {dx}},
\label{flux3}
\end{equation}
which leads to the well known generalization of Fick's law
for a space dependent diffusion coefficient\cite{schnitzer}.

To confirm these results, we performed simulations of Brownian walkers
following algorithm (3). We found that only the $\alpha=1$ case leads
to a uniform distribution of particles with no net flux through any
given surface while, at the same time, the average {\it individual}
positions exhibit a drift in the direction of the diffusion
coefficient gradient according to Eq(4). This situation of ``drift
without flux'' may be compared to the equilibrium situation of Brownian
particles subjected to an external force, such as their weight: If one
follows the motion of individual particles, an average downwards drift
is observed; however, there is no net flux because of the vertical
concentration gradient. In our isothermal case, the drift of
individual particles from lower to larger D(x) region does not lead to
a net flux because particles in the larger D(x) region diffuse further
than particles in the lower D(x) region. This physical situation
imposes the choice of $\alpha=1$ in algorithm (3), so that a particle
coming from a low $D(x)$ region makes a right step just equal to the
left step of that particle coming from a high $D(x)$ region and
arriving at the same point (see left part of Fig. 1).  

\begin{figure}
\psfig{file=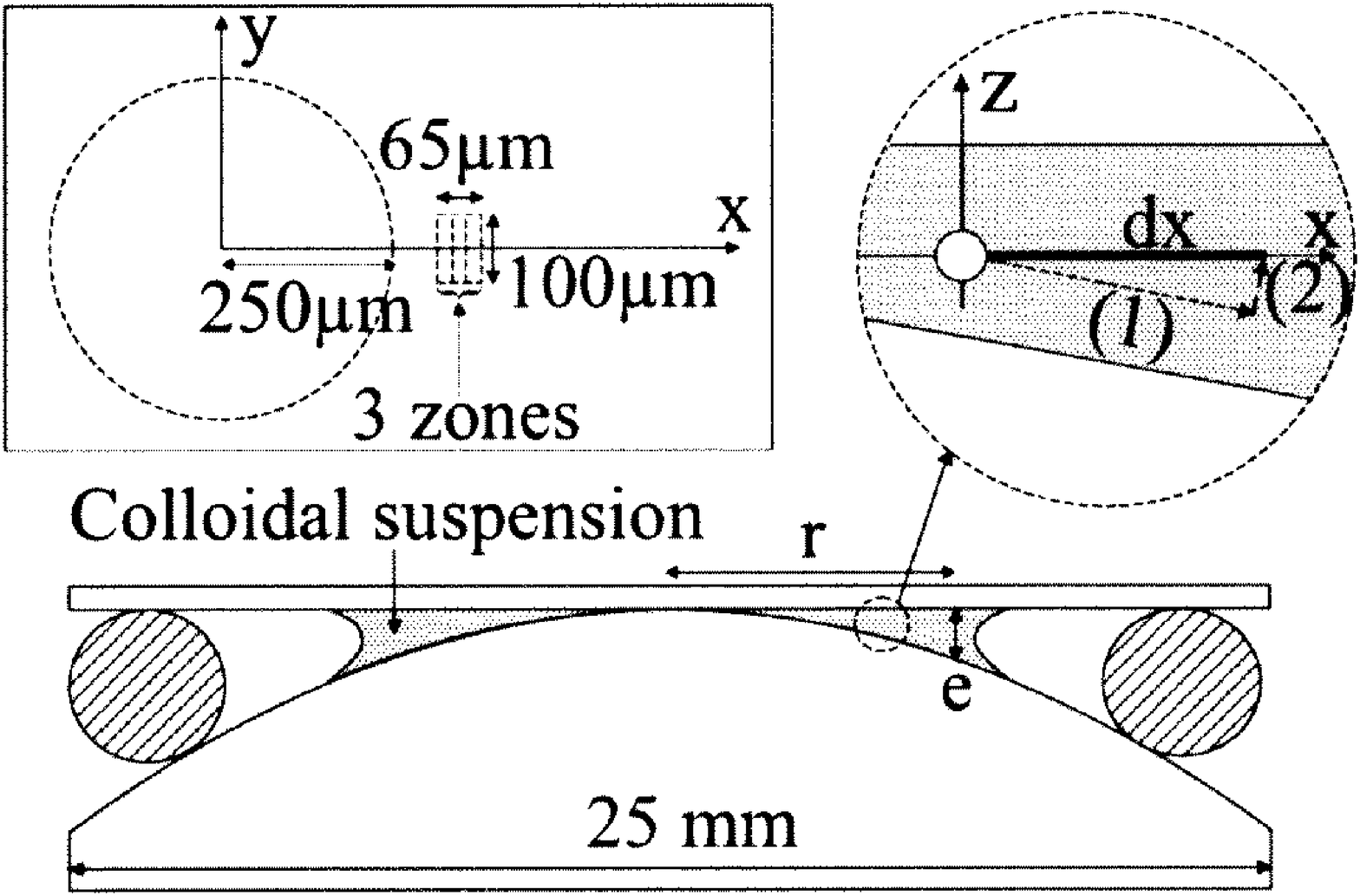,height=2.25in,width=3.5in}
\vskip-00mm
\caption{ 
Experimental set-up.  Rectangular inset is an enlarged top view
showing the center of the cell with the circular excluded volume and the
observation frame. Round inset explains the two contributions to the change
in diffusion coefficient when a particle moves a distance dx.
}
\end{figure} 

We have set up an experiment to check this drift without flux
prediction where particles, observed under a microscope, undergo
Brownian motion in a confined geometry.  When the confinement, $e$, is
of the order of the particle size, $a$, the diffusion coefficient
strongly depends on the value of $e/a$.  As the confinement was
position dependent, we observed Brownian walkers in an isothermal
situation but with a spatially dependent $D(x)$ due to a purely
hydrodynamic effect.

Polystyrene spheres, of radius $a=1\mu$m, were suspended in a mixture
of $H_2O+D_2O$ so as to cancel any sedimentation effects. Addition of
a surfactant ($2.2g/l$ of SDS) helped minimize particle aggregation or
adhesion to the walls. A drop of this mixture was placed between a
flat disk and a planar convex lens, Fig. (2), of curvature radius $R=
15.5$mm separated by an elastic O-ring. The flat and convex surfaces
were then brought into contact at the center of the cell by gently
squeezing the elastic joint, the remaining air providing the necessary
sample compressibility. The spacing, $e$, between the flat and curved
wall depends on the distance $r$ from the center of the cell as
$e=r^2/2R$. The contact between the two walls as well as the
dependence of $e$ on $r$ were carefully measured by monitoring the
Newton rings observed under the microscope. We used as a light source
a new super-radiant diode\cite{private} whose coherence length is less than
$100\mu$m.  This was important as this coherence length was long enough
to observe the desired Newton rings, but short enough to avoid any
other interference patterns due to all the cell interfaces, which were
visible with an ordinary diode laser and which completely masked the
relevant signal. The horizontal Brownian motion of the polystyrene
balls was observed through a microscope equipped with a long range
objective of magnification $50X$, followed by a CCD camera coupled to
the microscope via an eyepiece of magnification $8X$. The video
signal was processed in real time by a computer, which recorded, every
$3$s the horizontal position, size and shape of all objects in a
rectangular frame $65\mu{\rm m}X100\mu$m.  This time interval was long
enough to allow the image analysis of all particles present in the
frame and small enough so that the particle's average displacement was
only a fraction of their diameter.  

As the particles' confinement, $e$, was related to their distance,
$r$, from the center of the cell, we were able to explore different
confinement regions by moving the observation frame in the horizontal
plane. The explored $e$ varied from $2.5\mu$m to $11\mu$m. The
vertical Brownian motion of the particles over this small vertical
range could not be monitored. However, we took that motion into
account when interpreting the data by averaging the particle's
vertical position over the confinement range.  The volume fraction of
polystyrene balls, of the order of $1\%$, was chosen so as to allow
the monitoring of a fairly large number of particles at the same time
(around $30$ for $e=3\mu$m) to improve the statistics in the data
analysis. Following the particles' positions from one frame to
another, the program analyzed a great number of trajectories (more
than $10^5$ for each run). When two particles got closer than twice
their diameter, the program treated them as ``dimers'', their
trajectories as ``monomers'' were ended at that time and were no longer
used to determine the diffusion coefficient or the drift. In that way,
the role of particle interactions, which have a range smaller than a
couple of particle diameters, could be safely ignored.

\begin{figure}
\psfig{file=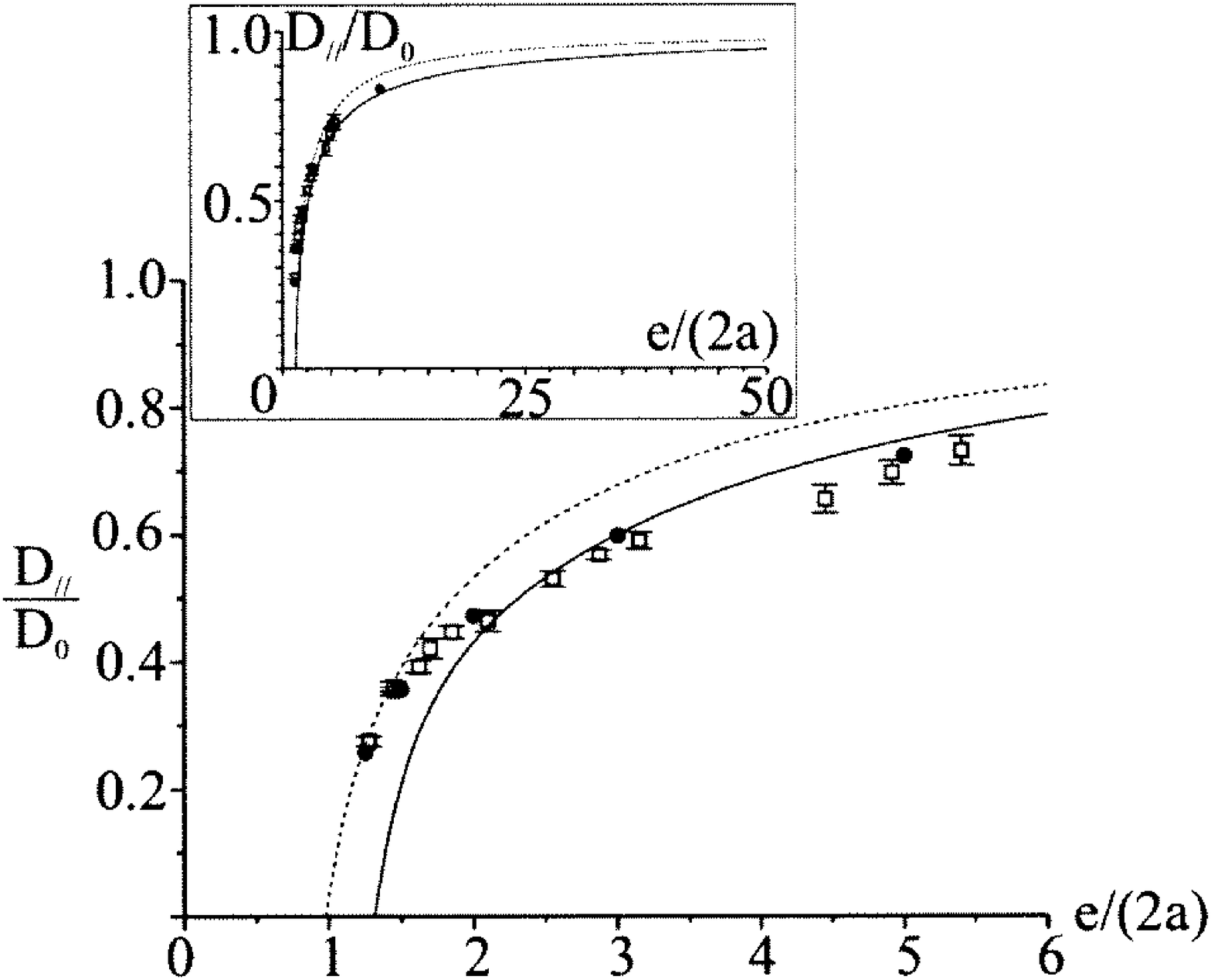,height=2.25in,width=3.3in}
\vskip-00mm
\caption{ 
$D_{\parallel}/D_0$ with respect to relative confinement e/2a . Open
squares are the experimental data, black dots were calculated by the
collocation method, and dotted and solid lines follow analytical
approximate solutions.
}
\end{figure}

The confinement dependence of the diffusion coefficient was determined
as follows. A given observation frame was divided into 3 zones (see
inset, Fig. (2)), each corresponding to an approximately constant $e$.
For each zone we averaged the particle's displacement squared, either
in the x, or in the y directions, as a function of time, and checked
that indeed they followed the usual diffusion law,
\begin{equation} 
\langle x^2\rangle = \langle y^2\rangle = 2D_{\parallel}(e) t.
\label{diffu}
\end{equation}

By moving the frame to different locations, we were able to explore a
range extending from $e/2a=1.2$ to $11$. The bulk diffusion constant,
$D_0$, was determined using the well-known relation $D_0=kT/6\pi \eta
a$, where $\eta=0.99X10^{-3}$SI is the viscosity of the water- heavy
water mixture, yielding $D_0=1.92X10^{-13}$m$^2$/s. The experimental
values for $D_{\parallel}/D_0$ are shown in Fig. (3) (white squares)
and fit remarkably well the available theoretical predictions (black
dots) using the collocation method\cite{ganatos} averaged over all the
possible vertical positions $z$ of the particle for a given $e$,
i.e. with $a\le z\le (e-a)$.  For comparison, we also plotted (solid
line) the analytical solution obtained using the Faxen
expression\cite{faxen} for the position dependent drag of a particle
moving parallel to a single wall, then adding the effect of each of
the two walls, and averaging over the vertical position $z$.  This
solution clearly overestimates the reduction of the diffusion
coefficient of a particle trapped between two parallel walls,
particularly as the relative confinement, $\epsilon= e/2a$ reaches its
lower limit 1.  We also plotted (dashed lines) in Fig. (3) the
analytical expression for the drag of a particle trapped just in the
middle of two parallel planes.\cite{faxen} The impossibility of
averaging over $z$ an expression only known for $z=e/2$ explains the
observed discrepancy which goes to zero as the relative confinement
approaches its limit, $1$, where the z-average becomes irrelevant.

To demonstrate the existence of an individual drift of the particles,
we fixed the center of the observation frame at a position $y=0$ and
$x=300\mu$m, corresponding to an average $e/2a=1.5$ so that all
particles present in the frame were outside the excluded volume
(i.e. $e\ge 2a$), and had a diffusion coefficient with the largest $x$
dependence, but no y dependence (to first order). For the
determination of $\langle x(t)-x(0)\rangle $ and $\langle
y(t)-y(0)\rangle $, each trajectory was divided into independent paths
lasting a time $t$, each contributing to the evaluation of the average
drift during time $t$. The results are shown in Fig. (4), and reveal a
drift in the Brownian walker position along the $x$ direction, and
none in the $y$ direction along which the diffusion coefficient may be
considered as constant. The statistics of the results clearly
deteriorates as time increases: After recording trajectories for
typically a dozen hours, more than a hundred thousand independent
segments contributed to the determination of the drift at short times,
whereas only up to a few thousand independent segments were left for
$t=200$s. This is due to the fairly high particle concentration which
lowers the lifetime of a ``monomer'' (time during which a particle
doesn't approach another one to within 2 particle diameters), but
which was chosen as a compromise to have good statistics at short
times while allowing us to follow each particle during a reasonable
time, fixed at $200$s. It should be pointed out that in order to avoid
any bias in the statistics, for a trajectory segment to be valid and
included in the statistics, the position of a walker at instant $t=0$
had to be inside a region $15\mu$m away from the edges of the
observation frame. This condition ensured that after diffusing for
$200$s, the walker had less than $0.5\%$ chance to have covered
$15\mu$m, and was thus still present in the observation frame. Failure
to impose this condition resulted in the observation of a spurious
drift, in the opposite direction, due to an artificial selection of
walkers because of the experimental boundary conditions (limits of the
observation frame).

To compare our experimental results with the theoretical predictions,
Eq. (4), we evaluated the diffusion gradient
encountered by the walkers present in the observation frame. It is
important to realize that as a walker moves a distance $dx$ (see round
inset in Fig. 2), its diffusion coefficient $D_{\parallel}$ varies first
because the confinement varies (path (1) parallel to the bottom wall,
i.e. at constant $z$), and second because at constant confinement, the
particle's height $z$ changes (path (2)). Adding both contributions
and averaging over the vertical position z of the walker yields:
\begin{equation} 
\big\langle {{dD_{\parallel}(e,z)}\over {dx}}\big\rangle_z = 
{x\over R}\big\langle
{{\partial D_{\parallel}(e,z)}\over {\partial e}}\big\rangle_z + {{x}\over
{2R}}\big\langle {{\partial D_{\parallel}(e,z)}\over {\partial
z}}\big\rangle_z.
\label{diffu2}
\end{equation}

Using our experimental data and results of the collocation method, we
found $\langle {{dD_{\parallel}(e,z)}/{dx}}\rangle_z\approx
2.2X10^{-9}$m/s. This value of the slope is used to plot the straight
dotted line on Fig. 4. The experimental data are thus in good
agreement with the predicted drift corresponding to the expected
$\alpha = 1$ value.

\begin{figure}
\psfig{file=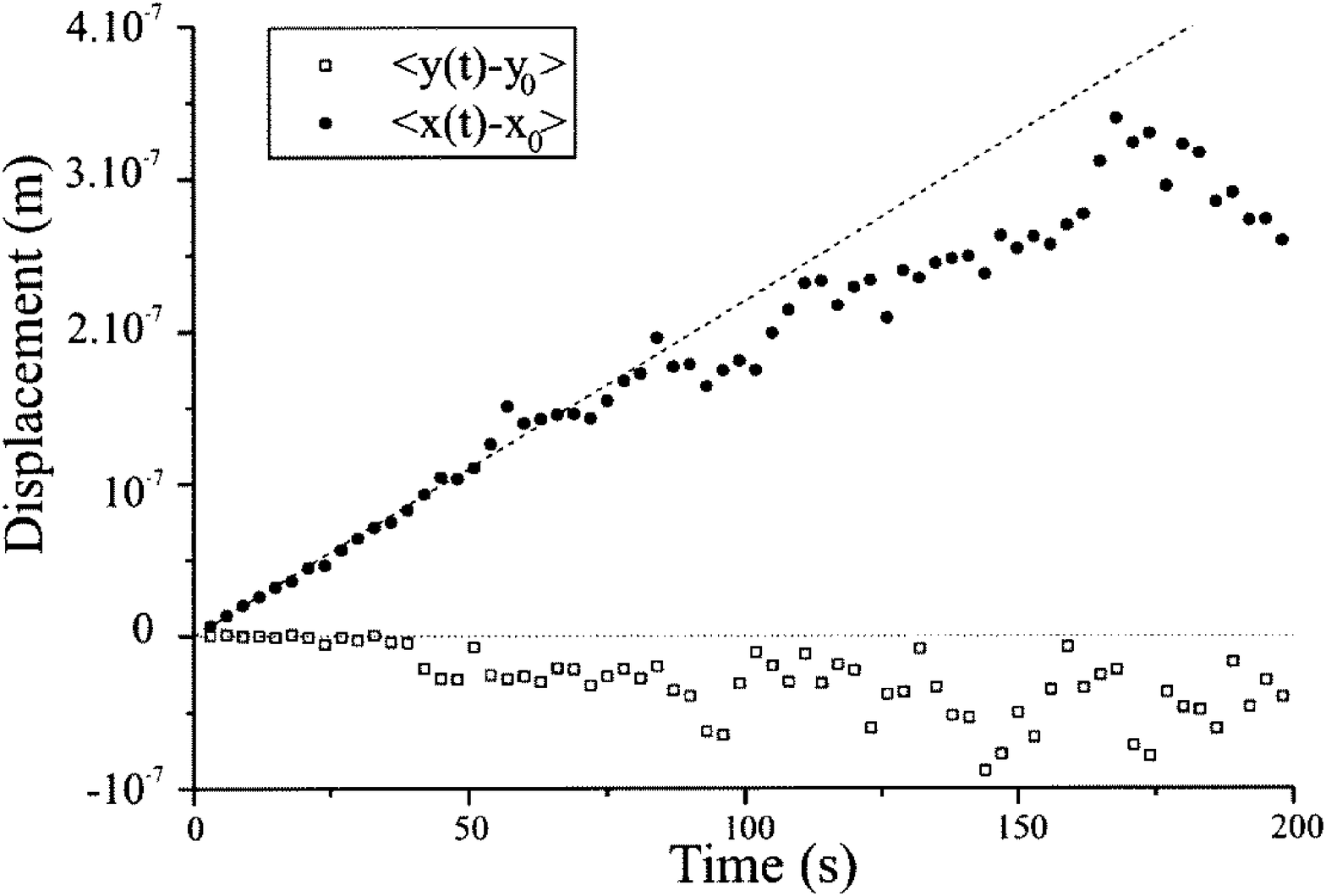,height=2.25in,width=3.5in}
\vskip-00mm
\caption{ 
Average position of the walkers as a function of time, along (black
dots) and perpendicular (open squares) to the diffusion gradient.
}
\end{figure}

Finally, to claim drift without flux, it is not sufficient only to
demonstrate the drift: we must also demonstrate the absence of
flux. If a flux was due to the observed drift $dD/dx$, we would expect
a radially outwards flux of $\rho dD/dx$ particles, which would empty
our observation screen in less than a day. Furthermore, if this flux
were to be balanced by a concentration gradient, one can show that a
concentration change by $30\%$ over a distance of $60\mu$m would be
necessary. Experimentally, we observed no flux and no concentration
gradient over a period of a week or more, which is consistent with the
Boltzmann requirement of a uniform concentration in the absence of a
temperature gradient.

\end{document}